# A Sensor-Based Simulation Method for Spatiotemporal Event Detection

Yuqin Jiang [1,2], Andrey A. Popov [3], Zhenlong Li [2,4,*], Michael E. Hodgson [2] and Binghu Huang [5]

1   Department of Geography and Environmental Studies, Texas State University, San Marcos, TX, 78666 USA
2   Geoinformation and Big Data Research Laboratory, Department of Geography,
University of South Carolina, Columbia, SC, 29208, USA
3   Oden Institute, The University of Texas at Austin, Austin, TX, 78712, USA
4   Geoinformation and Big Data Research Laboratory, Department of Geography, The Pennsylvania State University, University Park, PA, 16802, USA
5   College of Oceanography and Space Informatics, China University of Petroleum (East China), 265800, China
*   Correspondence: zhenlong@psu.edu

**Abstract:** Human movements in urban areas are essential to understand human–environment interactions. However, activities and associated movements are full of uncertainties due to the complexity of a city. In this paper, we propose a novel sensor-based approach for spatiotemporal event detection based on the Discrete Empirical Interpolation Method. Specifically, we first identify the key locations, defined as "sensors", which have the strongest correlation with the whole dataset. We then simulate a regular uneventful scenario with the observation data points from those key locations. By comparing the simulated and observation scenarios, events are extracted both spatially and temporally. We apply this method in New York City with taxi trip record data. Results show that this method is effective in detecting when and where events occur.

## 1. Introduction

Human movement within urban areas serves as a vital indicator of a city's functionality and the interaction between its inhabitants and the physical environment [1,2]. Understanding human mobility patterns in a city is essential for urban planning [3,4], transportation [5,6], and emergency management [7,8]. The increasing availability of data on human mobility creates remarkable opportunities for research. These data are derived from diverse sources including cell phone records [9–11], geolocated social media posts [12,13], and travel records [14,15]. Researchers across multiple disciplines are leveraging these data to enhance the modeling and understanding of both the regularities and anomalies in human mobility patterns, contributing significantly to our comprehension of urban dynamics.

Activities in the city are inherently complex and full of uncertainties. Events or anomalies are hard to define due to the complexity and scale of a city. Some events, like sports games and festival parades, are scheduled and foreseeable, allowing for advanced planning and preparation for impacts such as road closures and large crowds [16,17]. Conversely, certain events like hurricanes or winter storms, while predictable in their occurrence, present uncertainties in their intensity and affected areas [8]. Some other events are unexpected, such as car accidents or blackouts [5]. Those unexpected events have various impact scales and may not even be recorded. Urban events or anomaly detection methods have been developed with multiple data sources, including trajectories [18–20], origin–destination (OD) trip records [15,21,22], social media posts [23–25], cell phone data [26,27], and videos/images from surveillance cameras [28–31].

In this paper, we propose a novel sensor-based method for event detection with spatiotemporal big data. The method will identify key locations in the data and simulate a regular uneventful scenario. These key locations are defined as "sensors", which have the strongest correlation with the whole dataset. We define the simulation to be an uneventful scenario estimated based on the observation from these sensors. The discrepancy between a simulation and the observed data defines the events in space and time. Specifically, we make use of the Discrete Empirical Interpolation Method (DEIM), which is an extension of the Principal

Component Analysis (PCA) used to decompose the data not only into its principal components in the data space but also its key spatial locations in the geographic space [32,33]. Although the DEIM has been applied in other fields, such as nuclear desalination plant [34], water distribution [35], and fluid flow reconstruction [36], this is the first application of the DEIM in spatiotemporal event detection. We apply this method to billions of taxi OD trip records in New York City (NYC) from 2009 to 2012. Results show that this method can first identify the most important locations as the sensors' location. Secondly, by simulating uneventful scenarios, and comparing with the true observation, this method can find out the most events day during the study timespan. In addition, it can display the spatial patterns of discrepancy to illustrate the spatial distribution of disruptions across the study area.

## 2. Related Work

Event detection or anomaly detection methods using human mobility data have been extensively studied and used in multiple areas, including city-scale event detection [37–40], traffic conditions [17,41], environment management [42–44], infectious diseases [13,45,46], and natural hazards [47–50]. These methods have been developed based on multiple data sources, including social media data [13,42,51], vehicle trajectory data [52–54], OD trip records [15,55,56], and cell phone data [26,57].

Given the diverse strengths and weaknesses of each data type, specific methods and applications are customized to maximize the utility of each source. Social media data, for instance, offers a wealth of information beyond geolocation, typically including texts, user profiles, images, or videos. Methods and tools have been developed to merge geolocation with topics retrieved from texts, images, or videos, to identify events or anomalies across space and time. For example, studies have applied topic modeling methods to identify the most dominant topics discussed on Twitter at a given space and time to identify events [58–60]. Furthermore, advancements in image recognition have improved event detection accuracy with social media data by merging text and image analyses [61,62]. Despite these advancements, social media data are limited by representativeness issues, with biases in user demographics and a scarcity of geotagged posts [63–65].

Cellphone or mobile phone data have a better population representation. However, the Call Detail Record (CDR) data, the widely used mobile phone data, is based on the locations of signal towers. It records the signal tower's location when a cell phone user is making a call or sending/receiving a text message. Therefore, the spatial precision is dependent on the density of signal towers [15,27,43]. Vehicle-based datasets, especially in urban contexts, offer higher precision. These datasets are primarily classified into trajectory data and origin–destination (OD) data. Trajectory data record the vehicle's location at a certain time interval, which can indicate the actual route; however, the status of the vehicle is unknown, such as whether a taxicab is occupied or empty [19,66,67]. OD data, on the other hand, only record a trip's origin and destination locations, omitting the actual driving routes. Although the exact driving routes remain unknown, OD data mark the demand for travel with vehicles across space and time [55,56]. Research utilizing OD taxi data has been useful in understanding city landscapes, optimizing taxi dispatching, and discerning individual travel patterns [1,15,68]. In this paper, we utilize taxi trip OD data, treating each trip's origin and destination as distinct points for analysis.

Traditional event detection methods include clustering and time-series analysis. Clustering, an unsupervised learning method, allows for the partitioning of a city into multiple functional regions to identify anomalies in human mobility patterns using OD trip data. For example, DBSCAN has been used to identify hotspots in OD trips as clusters for pick-up and drop-off activities [55,68]. DBSCAN clusters can also identify activities with social media check in data [69]. K-means clustering algorithm has been used to partition a city into multiple regions based on taxi pick-up and drop-off activities [70]. The K-means clustering method has also been used to identify regularities in individual's mobility pattern and, thus, can help to identify anomalies from outliers of clusters [71]. Spatial Scan Statistic and its extensions are also used to identify spatiotemporal clusters by comparing observed distribution with Monte Carlo simulated distributions [72]. In another approach, Latent Dirichlet Allocation is used on cubes constructed from origin and destination trip records to identify clusters [2]. Network-based clustering methods have also been applied to significant OD flow detection, particularly with shared bike data [73]. Nonetheless, these clustering techniques primarily focus on spatial distribution patterns at specific observation times, potentially overlooking long-term trends [15]. To address this limitation, time-series analysis methods have been introduced to capture temporal trends. Discrete Fourier Transformation, for instance, is employed to discern periodicity in human mobility patterns, distinguishing regular daily and weekly changes from anomalies [40]. To further dissect long-term and seasonal trends, seasonal and trend decomposition methods are used to break down time-series patterns into long-term trends, seasonal periodicity, and residuals, with significant residuals flagged as events [15]. However, time-series analysis requires data to be sorted chronologically, which can be computationally intensive for large datasets.

In certain disciplines, when the complete observation dataset is not available, event or anomaly detection often relies on simulation with limited observation data points. These simulation-based methods become critical when data can only be obtained through physical sensors in the field. For instance, hydrological models use data from gauges to simulate the probability of flash flooding and issue warnings accordingly [74,75]. Similarly, numerical weather prediction models depend on data assimilation methods, wherein forecasts are augmented with sparse, noisy atmospheric observations within a Bayesian framework [76,77]. In addition, a generalized machine learning framework was developed for unsupervised, high-performance, spatiotemporal event detection. This approach first builds reduced-order representations of spatially local information, followed by the application of a discrepancy metric to discern the occurrence and location of events [78].

In this paper, we introduce a novel approach for event or anomaly detection utilizing the Discrete Empirical Interpolation Method (DEIM) [32,33], a simulation method rooted in Principal Component Analysis (PCA). PCA transforms an original dataset into dominant orthogonal components, thereby revealing correlations between predictive and observation variables. Based on this concept, PCA-related algorithms have been developed to identify the most dominant factors in traffic patterns, and thus, anomalies can be detected [79–81]. The DEIM further extends PCA by identifying not only the most dominant linear combination of variables but also the most dominant variable in each component. The DEIM has been used in areas including nuclear desalination plant [34], water distribution [35], and fluid flow reconstruction [36].

## 3. Methodology

### 3.1. Method Overview

This study proposes an event detection method with optimal sensor placement and interpolation using the DEIM. This method consists of four steps. The first step is data preparation. In this step, we divide the study area into spatial cells and count travel activities in each cell. In the second step, we determine the optimal number of sensors and their locations. In this study, a sensor is an abstract concept such that only the true observation data at the sensor location is treated as known data for the simulation. The third step is to simulate a regular uneventful scenario based on the observation data at the sensors' locations. Given a time period of interest, we obtain the observation data only at the sensors' locations. Then, for each temporal unit, a simulation is generated based on these sensors' observations. This simulation is defined as the uneventful situation that we use to compare with the observations. The last step is to compare the simulations with the observations. At the aggregated level, we calculate Root-Mean-Squared Error (*RMSE*) to identify which temporal unit has the largest event. Differences between the simulation and the observation at the cell level can be mapped for the spatiotemporal distribution of the events. Figure 1 shows the overall workflow in this study.

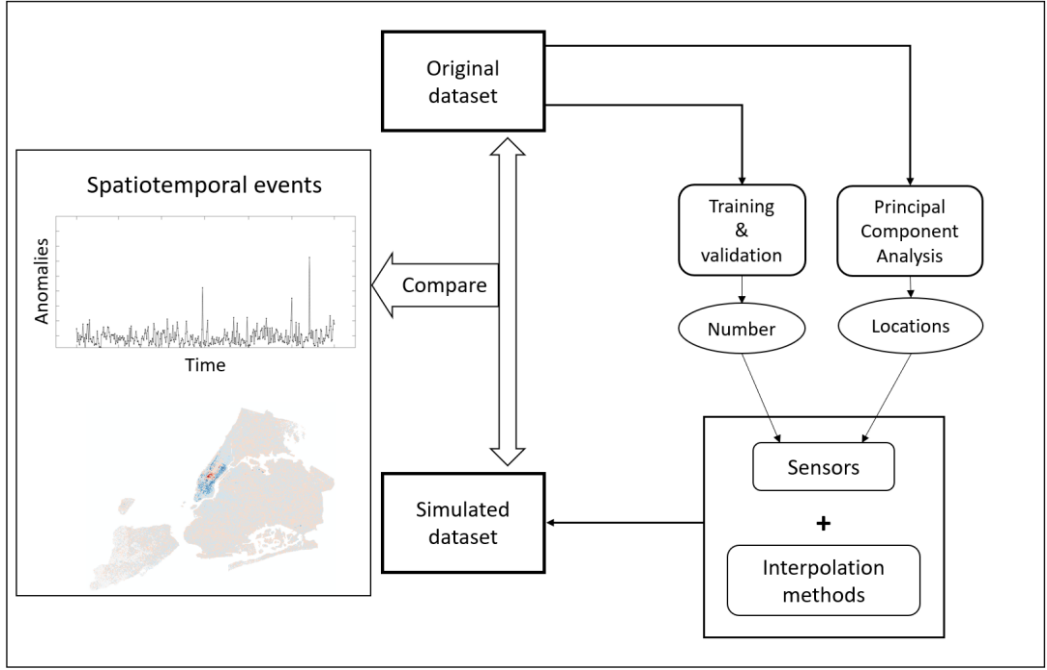

**Figure 1.** Illustration of the overall workflow.

## 3.2. Model Human Mobility Data

We partitioned the study area into small spatial units, commonly square- or rectangle-shaped cells. Each cell is assigned a unique identification number. Then, for each temporal unit (e.g., a day or an hour), we summarize the human mobility signals in each cell. Such signals can be the number of taxi pick-ups or drop-offs, the number of geotagged social media posts, or cell phone signals. By mapping human mobility data per cell, we can generate a cell-based two-dimensional map for one temporal unit. By creating and overlaying such maps for all the temporal units, we create a space–time cube from human mobility data (Figure 2a). In this space–time cube, X and Y correspond to the geographic location of the given cell in the two-dimensional map. In this example, there are $n$ rows and $m$ columns in the two-dimensional map, and thus, there are a total of $n \times m$ cells for each temporal unit. The depth of this cube, $T$, represents the number of temporal units.

We then transform this space–time cube into a matrix **A** (Figure 2b). This matrix has $k$ columns, representing $k$ temporal units. The number of rows for this matrix is $n \times m$, representing a vectorized two-dimensional map. The row identification number corresponds to the unique cell identification number in the study area. This matrix **A** is used for the subsequent analysis.

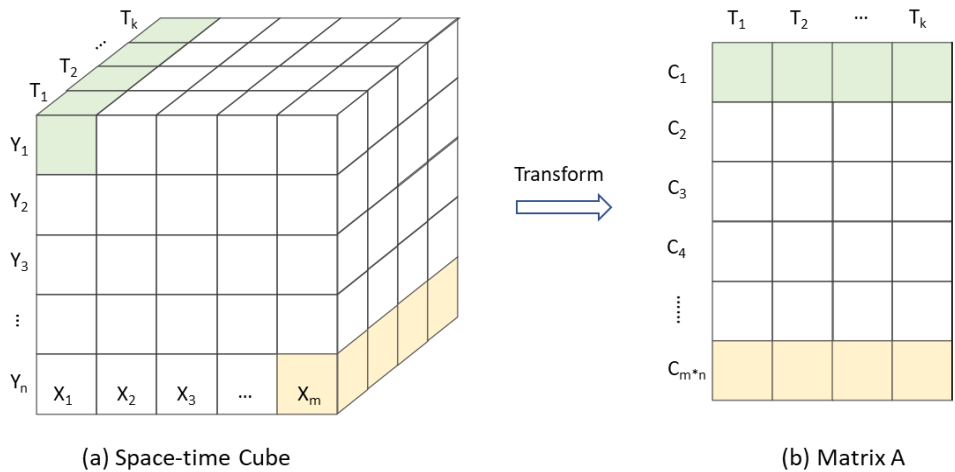

**Figure 2.** Modeling the spatiotemporal human mobility data: (**a**) space–time cube in the geographic space summarizing the human mobility data and (**b**) matrix A transformed using the space–time cube.

## 3.3. Determine the Locations of Sensors

Locations of sensors are determined by PCA. PCA is a statistical technique to simplify and reduce the dimensionality of data while preserving their essential structures and patterns. It linearly transforms the original dataset into a new coordinate system, where the new directions (axes) contain the most dominant variances in the original dataset. These new directions are represented by principal components. The first principal component captures the maximum amount of variation in the original data, and each subsequent component captures the remaining variations in the decreasing order of importance [82–84]. PCA is robust to handle raw data where the number of observations (columns) is larger than the number of variables (rows) [85,86]. When the input dataset has more variables than observations, PCA generates a spurious correlation, which does not reflect the true relationship. Unfortunately, this is the common case in spatio-temporal analysis as we usually have more spatial units (rows) than temporal units (columns). To solve this problem, the DEIM further extends PCA by not only identifying the dominant vector but also identifying the dominant variables. In PCA, each component takes all the spatial units, which means each component needs data from all the spatial units. The DEIM takes the first few dominant components and identifies the most important spatial unit in each component. The goal of the DEIM is to find the most dominant spatial location that indicates the strongest correlation. Specifically, the DEIM sequentially looks at each of the dominant components from PCA and finds the variable with the largest coefficient.

In this demonstrative example, the original dataset matrix has $k$ temporal units ($k$ columns). Let $x$ be the total spatial unit, where $x = m \times n$ from Figure 2b. Therefore, the component matrix resulted from PCA has $k$ components in total. The first $q$ components explain 96% of the variance and the rest $(k - q)$ components explain 4% of the variance (Figure 3a).

Let matrix $\mathbf{U}$ be the first $q$ component (the blue rectangle). The size of $\mathbf{U}$ is $x$ by $q$, where $x$ is the number of spatial units and $q$ is the number of the dominant components we decide to use. Since the number of spatial units is typically significantly larger than the number of components, matrix $\mathbf{U}$ is a long and narrow matrix. The transpose matrix of $\mathbf{U}$ is $\mathbf{U}^T$, which is a wide and short matrix.

Because $\mathbf{U}$ is the matrix for dominant components, it is an orthonormal matrix, and so is $\mathbf{U}^T$. Therefore, $\mathbf{U}\mathbf{U}^T$ removes all the non-dominant components from the PCA result. $\mathbf{U}\mathbf{U}^T$ is nearly equal to an identity matrix but is not exactly the same. Because $\mathbf{U}\mathbf{U}^T$ removes all the non-dominant components, $\mathbf{U}\mathbf{U}^T \mathbf{A}$ creates an approximation of the original dataset $\mathbf{A}$.

$$\mathbf{A} \approx \mathbf{U}\mathbf{U}^T \mathbf{A} \tag{1}$$

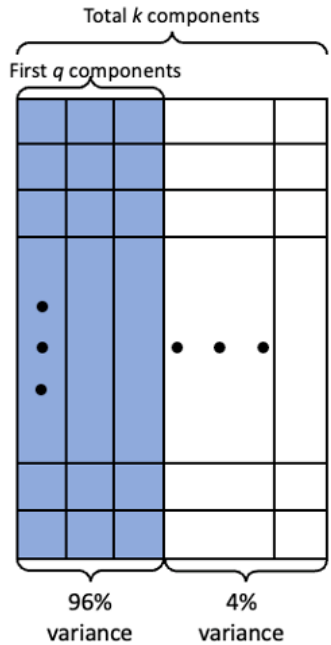
(a) Matrix **U** from PCA output

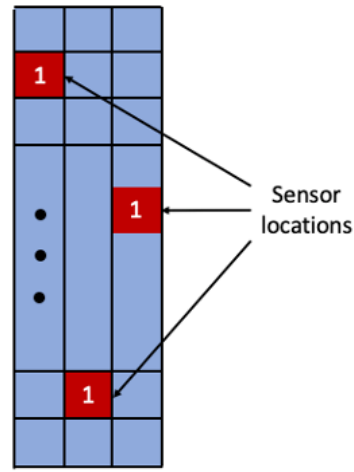
(b) Matrix **P** for sensors locations

**Figure 3.** Determination of sensor locations.

In this example, matrix **U** is for the first $q$ components. Therefore, the size of **U** is $x \times q$. In each component (each column of **U**), we find out the element has the largest absolute value. The first dominant component represents the first sensor, and the second component represents the second sensor. This process continues until we reach the pre-determined number of sensors. In this example, we will have $q$ sensors. For example, the $q$ red squares (spatial units) indicate locations for the $q$ sensors (Figure 3).

After we determine the locations of the $q$ sensors, we can define a matrix **P**, representing the sensors' locations. In **P**, cells in which a sensor is located have the value one, and all other cells have the value zero. The size of the matrix **P** is $x \times q$, same as **U**, but all elements of **P** are zero, except elements at the $q$ locations of sensors have the value one (Figure 3b). Because **P** is also orthonormal, $\mathbf{P}\mathbf{P}^T$ is nearly equal to an identity matrix.

*3.4. Determine the Optimal Number of Sensors*

In this event detection model, the only parameter is the number of sensors. To determine the optimal number of sensors for use, we first train the model using the dataset in which we are interested. However, because this method is based on PCA, the more sensors are used, the less unexplained variance occur, which results in an overfitted model. To mitigate this issue, we use an external validation dataset to identify an optimal number of sensors for use. The validation dataset is very similar to the training dataset. For example, the validation data can be from the same study area but during a different time period.

We first apply this model to the training dataset with different number of sensors. Each run uses a different number of sensors and thus generates a corresponding *RMSE*. Let $a_{ij}$ be the element from **A** at $i$th row and $j$th column and $\tilde{a}_{ij}$ be the element from $\tilde{\mathbf{A}}$ at $i$th row and $j$th column. We calculate the *RMSE* from the difference between the observed dataset **A** and the simulated data $\tilde{\mathbf{A}}$ using the RMSE calculation method:

$$RMSE = \sqrt{\frac{\sum(a_{ij} - \tilde{a}_{ij})^2}{n}} \qquad (2)$$

where $n$ is the number of cells in the study area. This *RMSE* represents how large the difference is between the observed dataset and the simulated dataset.

This *RMSE* is a metric for the overall simulation error for the dataset. Therefore, the training dataset and the validation dataset do not need the same number of temporal units. In the validation round, we apply the locations of sensors to the validation dataset for simulation. The curve for models with the training dataset continues to decrease as the number of sensors increases, but the *RMSE* for the validation dataset reaches its lowest point. We use the number of sensors corresponding to the minimum *RMSE* point as the optimal number of sensors. This workflow is shown in Figure 4.

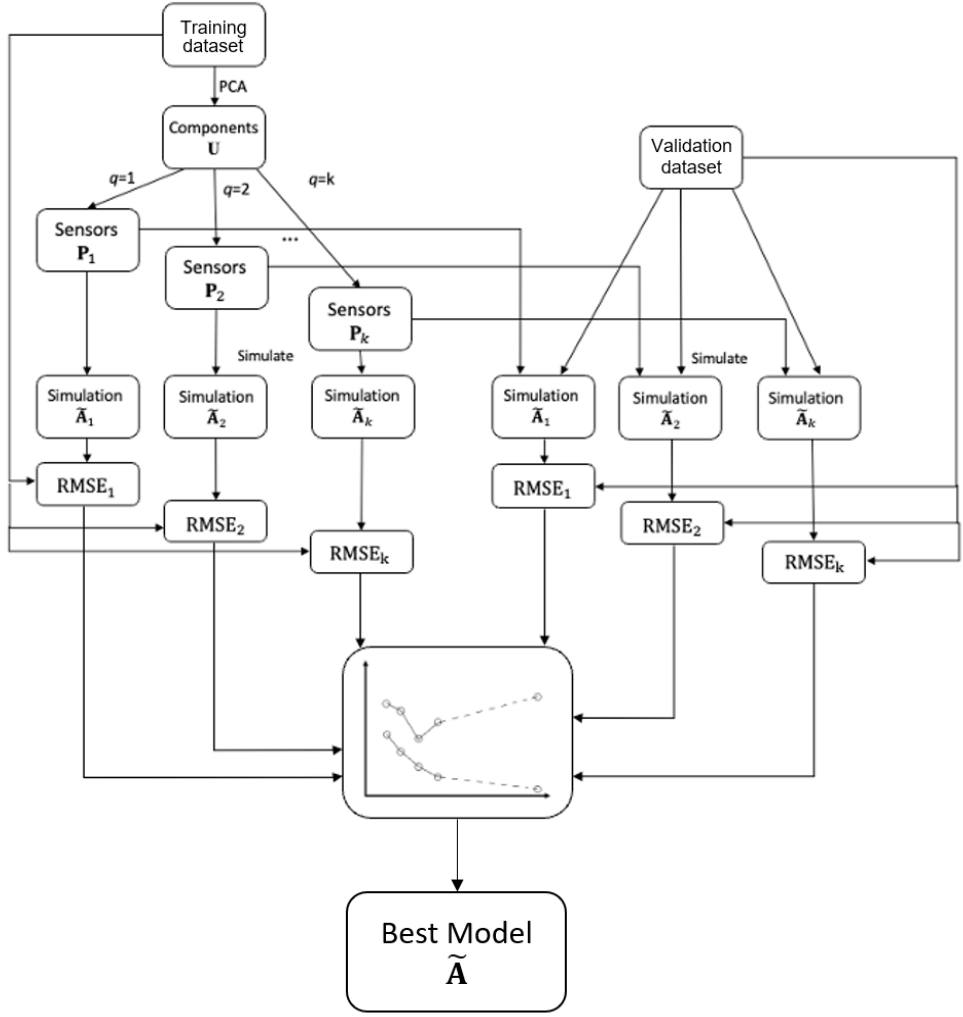

**Figure 4.** Workflow for identifying the optimal number of sensors and their locations.

*3.5. Simulate the Uneventful Scenario*

A traditional sensor is a physical equipment applied in the field to collect data (e.g., temperature and humidity). Based on the discrete data collected from sensors, the whole dataset is simulated. In our application, we treat sensors as the spatial unit with a known data point in all the temporal units to simulate a complete dataset. As we only use $q$ sensors from the first $q$ components, such simulation is an approximation of the original dataset. For example, in Figure 3a, the first $q$ components explain 96% of the variance of the dataset, this simulated uneventful situation ($\tilde{\mathbf{A}}$) is about 96% time equal to the original dataset ($\mathbf{A}$), and the other 4% of the difference are what we are interested as events.

Since the only known data are at the sensors' locations, these known data can be written as $\mathbf{P}^T \mathbf{A}$. Given Equation (1), we can obtain Equation (3):

$$\mathbf{P}^T \mathbf{A} \approx \mathbf{P}^T \mathbf{U} \mathbf{U}^T \mathbf{A} \tag{3}$$

Then, we divide $\mathbf{P}^T \mathbf{U}$ from both sides of Equation (3), and we obtain Equation 4:

$$(\mathbf{P}^T \mathbf{U})^{-1} \mathbf{P}^T \mathbf{A} \approx \mathbf{U}^T \mathbf{A} \tag{4}$$

Then, we multiple $\mathbf{U}$ on both sides of Equation (4), and we obtain Equation 5:

$$\mathbf{U}(\mathbf{P}^T \mathbf{U})^{-1} \mathbf{P}^T \mathbf{A} \approx \mathbf{U} \mathbf{U}^T \mathbf{A} \tag{5}$$

Because $\mathbf{U} \mathbf{U}^T \mathbf{A}$ is an approximation of the original dataset $\mathbf{A}$ (Equation (1)), we can rewrite Equation (5) as Equation (6):

$$\widetilde{\mathbf{A}} \approx \mathbf{U}(\mathbf{P}^T \mathbf{U})^{-1}\mathbf{P}^T \mathbf{A} \qquad (6)$$

Equation (6) is the equation used to simulate the scenario without events.

*3.6. Detect Events*

The simulated uneventful scenario $\widetilde{\mathbf{A}}$ has the same size as the original dataset $\mathbf{A}$. Specially, each row and column match the corresponding spatial and temporal units. We first calculate the difference between the observation and the simulation for each spatial unit and temporal unit using Equation 2, where $a_{ij}$ is considered the element from $\mathbf{A}$ at *i*th row and *j*th column and $\tilde{a}_{ij}$ is considered the element from $\widetilde{\mathbf{A}}$ at *i*th row and *j*th column. We calculate the *RMSE* to compute the difference between the observed dataset $\mathbf{A}$ and the simulated data $\widetilde{\mathbf{A}}$ using the *RMSE* calculation method (Equation (2)), where *n* is the number of cells in the study area. This *RMSE* represents how large the difference is between the observed dataset and the simulated dataset.

For a given temporal unit, we pair up the observed travel demand and the simulated travel demand and compare the difference. We define the difference as an "Event Index" by using the observation travel demand number minus the simulated travel demand number (Equation (7)):

$$Event\ Index = observation - simulation \qquad (7)$$

We then map the Event Index for each spatial unit to identify the spatial distribution of events. When the Event Index is negative, it means that the observed travel demand is smaller than the simulated, as there are unexpected events that cause the travel demand to decrease. On the other hand, when the Event Index is positive, it means that more than expected trips are observed. In general, the absolute value of the Event Index indicates the magnitude of an event. For example, if Location A has an Event Index of 1000 and Location B has an Event Index of 10,000, it means that the travel demand at Location B is more impacted.

## 4. Case Study

New York City (NYC) consists of five boroughs: Brooklyn, Queens, Manhattan, the Bronx, and Staten Island. Based on the 2010 Census data, NYC has more than eight million residents living in about 800 km$^2$, which makes NYC have the highest population density in the United States. Among these five boroughs, Manhattan has the largest population. Manhattan has more than 2.5 million residents, making up about 30% of the NYC population.

Due to its high population density, residents in NYC have low private car ownership. The overall car ownership in NYC is 45% but only 22% in Manhattan. In addition, only 8% of Manhattan residents drive to work. Taxis, subways, buses, and recently ridesharing play essential roles in New Yorkers' daily mobility.

This study uses taxi trip records from 2009 in the NYC area provided by the New York City Taxi & Limousine Commission (NYC TLC), the major taxi company operating the famous yellow taxicabs in NYC. Based on the data from NYC TLC, a total of more than 143 million taxi trips were completed in 2009. For each taxi trip, the location (latitude and longitude) and time for pick-up and drop-off were recorded, but the route between the pick-up and drop-off was not included. Since we only use the pick-up and drop-off locations in this study, all other trip information was not included in analysis and discussion. Figure 5 shows the travel demand by taxi in NYC on 1 January 2009.

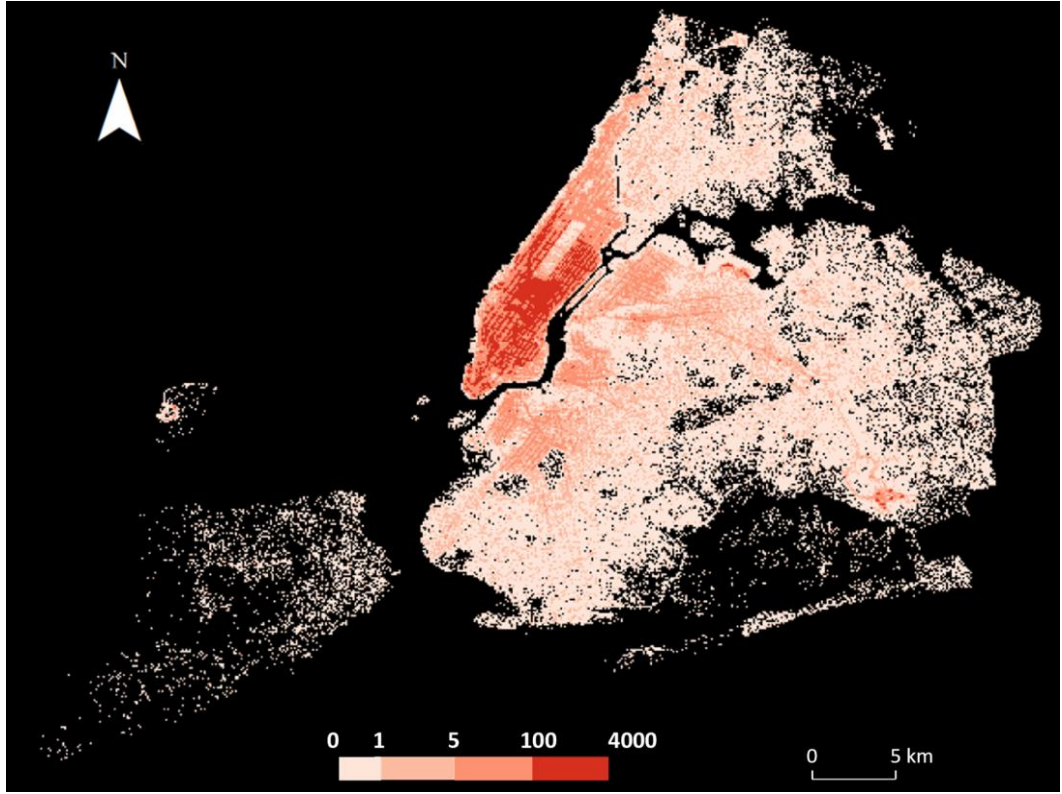

**Figure 5.** Travel demand by taxi in NYC on 1 January 2009.

We divided the study area into cells with a cell size of 110 m (0.001-degree latitude) by 80 m (0.001-degree of longitude). This cell size ensures that we have enough travel records to train and validate this model while not including too many intersections or important locations in each cell. NYC is represented by 87,838 cells and each cell is assigned with a unique identification number. Then, for each day, we count the taxi 'demands' for each cell as the summary of taxi trips starting from and ending in the given cell. We consider both pick-ups and drop-offs as demands for traveling and therefore both are treated the same. For example, for a given cell in a given day, if there were 100 pick-ups and 200 drop-offs, we consider this as 300 traveling demands. After this step, all taxi travel data are organized as a matrix **A**. The entry $a_{ij}$ means the travel demands for the $i$th cell on the $j$th day.

## 5. Results and Discussion

To validate the performance of this method, we apply this method to 2009 and 2012 NYC taxi data. For the 2009 model, we train our model on 2009 data and validate this model using a combination of 2010–2012 data. For the 2012 model, the training dataset is 2012 data, and the validation dataset is a combination of 2009–2011. We first generate models with different numbers of sensors and plot the overall *RMSE* for each model. Figure 6 shows the *RMSE* plotted for the corresponding number of sensors. For training datasets, the general trend is that *RMSE* decreases as the number of sensors increases. However, for validation datasets, the *RMSE* reaches its lowest point. This lowest point is the optimal number of sensors that we use in the model. Figure 6 shows the training and validation *RMSEs* for 2009 (left panel) and 2012 (right panel). Based on these two figures, the optimal number of sensors for 2009 is 13 and for 2012 is 8.

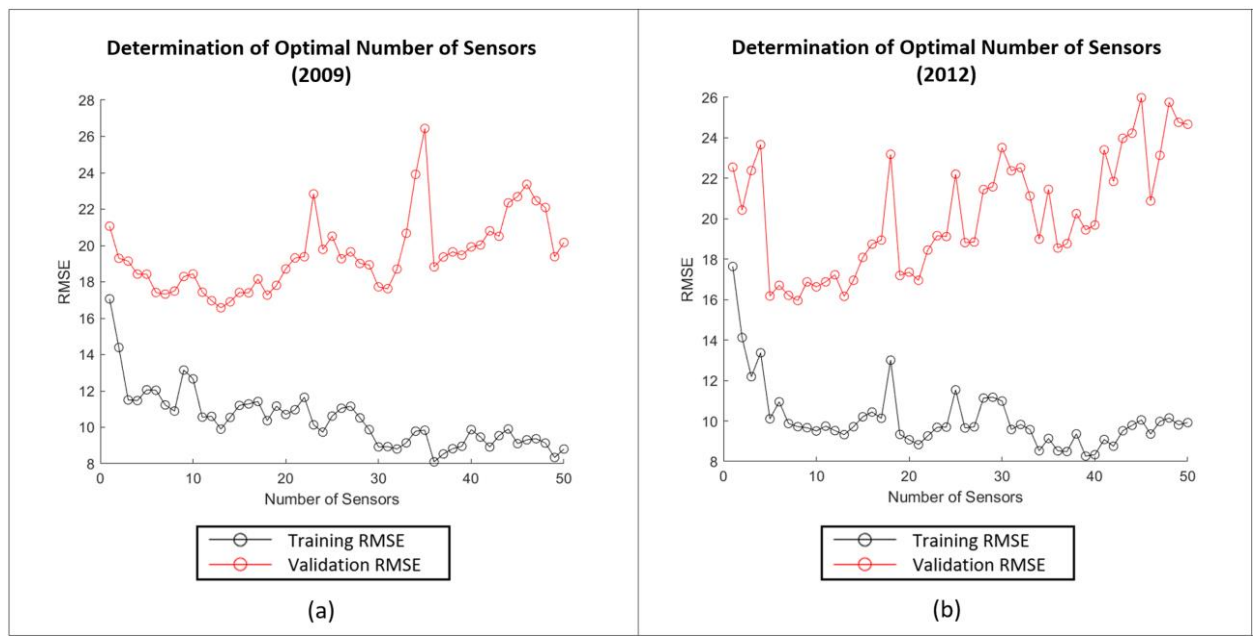

**Figure 6.** Determination of the optimal number of sensors for 2009 (**a**) and 2012 (**b**).

*5.1. Spatiotemporal Events in 2009*

Table 1 and Figure 7 illustrate the locations of the 13 sensors for 2009. The most important is Penn Station, which is the main transportation hub in Manhattan. Other sensors' locations include some important intersections in Manhattan, multiple locations near the LGA airport, and other landmarks in Manhattan. Noticeably, other than three sensors located in LGA airport, the other 10 sensors are all in Manhattan, indicating that Manhattan travel demands are the dominant patterns to analyze NYC.

**Table 1.** Locations of the 13 sensors for 2009.

| Sensor Number | Location |
|---|---|
| 1 | Penn Station |
| 2 | Intersection of Essex St. and Rivington St. |
| 3 | 1st Ave outside Tisch Hospital/NYU Langone Hospital |
| 4 | Intersection of W 42nd St. and 8th Ave. near Port Authority Bus Terminal |
| 5 | Javits Center |
| 6 | LGA drop-off area |
| 7 | LGA drop-off area |
| 8 | Lincoln Center for the Performing Arts |
| 9 | LGA taxi pick-up area |
| 10 | Madison Square Garden |
| 11 | 9th Ave between W 13th St and W 12th St |
| 12 | Empire State Building |
| 13 | 9th Ave and W 14th St. |

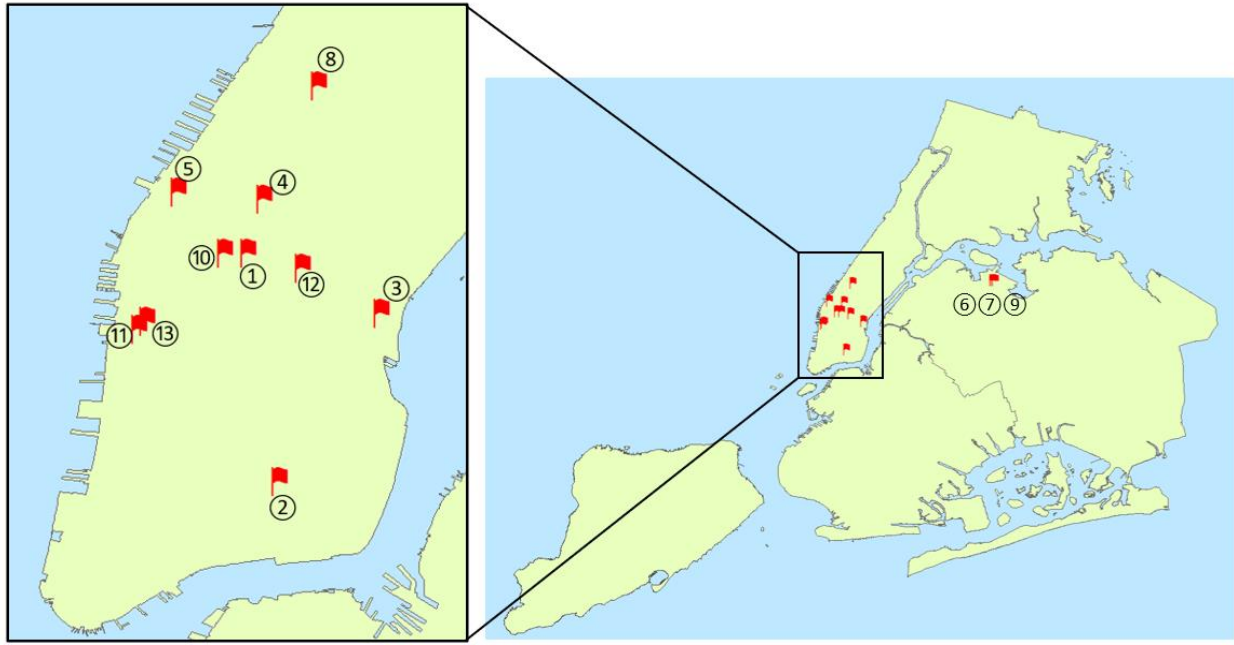

**Figure 7.** Locations of 13 sensors in NYC for 2009 event detection.

We then take the simulated scenario $\tilde{\mathbf{A}}$ using 13 sensors to calculate daily *RMSE*, which can identify the daily events over the entire 2009 year. The simulated matrix $\tilde{\mathbf{A}}$ is organized in the same way as the original dataset **A**, in which each column represents one day's travel demand of all cells. Therefore, calculating *RMSE* column-wise is to compare daily difference between the observed dataset and the simulated scenario. The results of daily *RMSE* are shown in Figure 8.

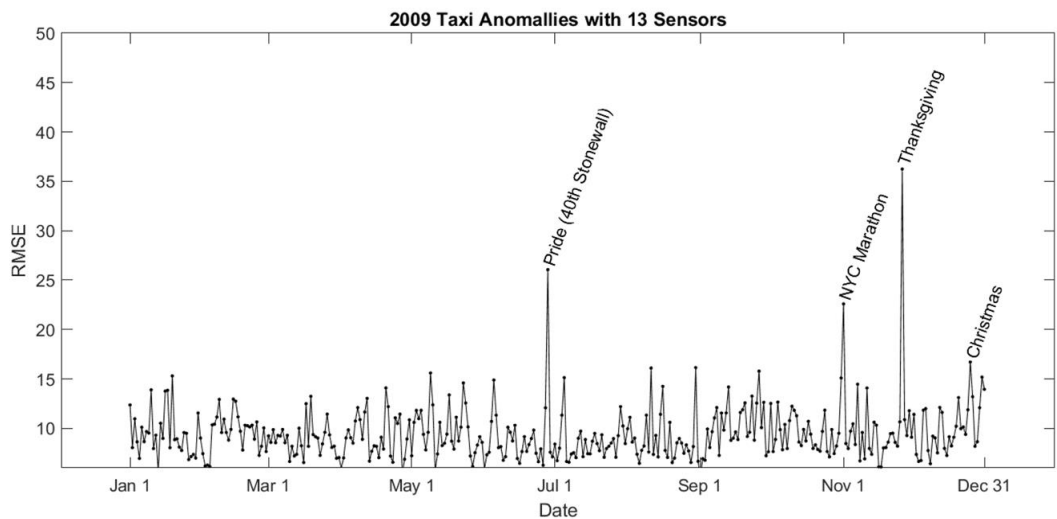

**Figure 8.** Daily *RMSE* showing the difference between observed travel demand vs. simulated uneventful scenario for 2009.

In 2009, the day with the largest difference between the observations and simulated uneventful scenario is 26 November, which is Thanksgiving Day. Other large differences are 28 June, the Pride March, 1 November, the day for NYC marathon, and 25 December, Christmas Day. As mentioned in Section 3.6, the Event Index for each cell indicates the difference between our model simulation and the actual observation. By mapping the Event Index, we can find the spatial events distribution.

Since most events in NYC are concentrated in Manhattan, as shown in Figure 9, in the results part, we only shown the Manhattan map for better illustrations of events distributions. We merge all the Event Index numbers into one vector and apply Jenks Natural Break classification [87] to determine the ranges of each category. Jenks Natural Breaks is a method used to classify numerical data into categories that minimizes the variance within each category while maximizing the variance between categories [88]. It is a popular method in cartography for choropleth maps, where the color classes provide obvious value categories in map [89,90]. In this way, the same color in all the maps represents the same Jenks category range, which is more convenient for comparison across different time periods.

Figure 9 shows the spatial distribution of the Event Index for 26 November, Thanksgiving Day of 2009. The two places with the largest positive Event Index are the Metropolitan Museum of Art and American Museum of Natural History, with 1770 and 1500 more observed trips than estimated, respectively. Midtown around Times Square experienced more travel demands on the Thanksgiving Day than estimated. However, most other places in Manhattan experienced less travel demand than estimated. The place with the largest negative Event Index is Port Authority, meaning less travel demands were observed than estimated.

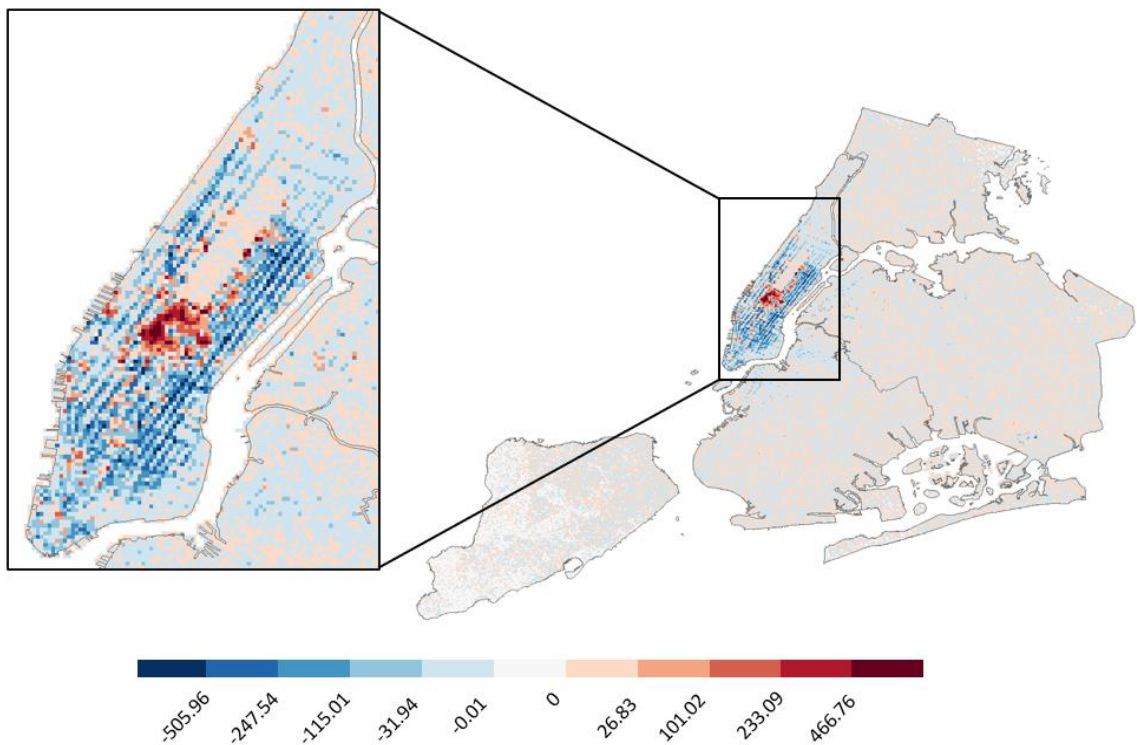

**Figure 9.** Spatial distribution of Event Index for Thanksgiving Day, 2009.

The day with the second largest overall *RMSE* is 28 June 2009, which is the day for NYC Pride March celebrating the LGBTQ community. Event Index distribution is shown in Figure 10a. 2009 is the 40th anniversary of the Stonewall Riots in NYC. Although the Metropolitan Museum of Art and American Museum of Natural History have positive Event Index, most of the cells with positive Event Index are clustered around Midtown Manhattan near Times Square. Most areas in the Upper East Side and the Upper West Side are in blue, meaning observed travel demands are less than estimated. Figure 10b shows the third largest *RMSE*, appeared on 1 November when NYC Marathon took place. This was the 40th annual marathon race in NYC. The end of this race was southeast of Central Park. Therefore, the travel demands were less than estimated due to road closure for the race. Also, the last part of the marathon route followed along the east side of Central Park. The blue area in the east part of the Central Park indicates the reduction of taxi travel demand due to the road closure. Cells with the largest Event Index are along the 2nd Avenue in the Upper East Side and areas at the south tip of Manhattan, near the Battery waterfront park, where the ferry to Statue of Liberty departures. This may be caused by tourism attracted to NYC for the Marathon events.

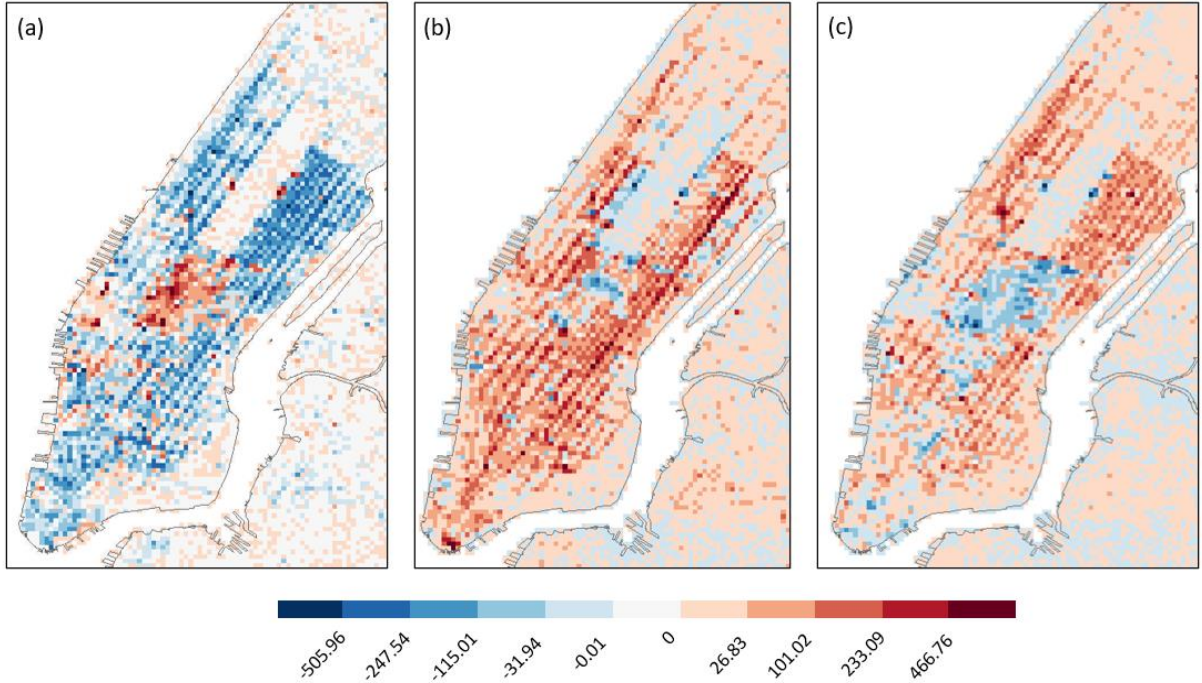

**Figure 10.** Event Index of Manhattan area in 2009 for (**a**) 28 June (Pride March celebrating the LGBTQ community), (**b**) 1 November (NYC Marathon), and (**c**) 25 December (Christmas).

The fourth largest *RMSE* appeared on 25 December, which is Christmas (Figure 10c shows the Event Index). The majority of Midtown Manhattan has a negative Event Index, meaning the number of actual trips is smaller than estimated. Midtown is the center of commerce. The decrease of taxi trips is caused by Christmas when most companies in Midtown were closed. Both the Upper East Side and the Upper West Side had more trips than estimated on Christmas. There are two cells with Event Index larger than 1000, meaning observed travel demands were more than estimated. One cell is located on the 2nd Ave., between E 31st St. and E 32nd St, and the other one is located on Broadway between W 67th St. and 68th St. Both cells are near some multi-floor residential condominium buildings. Such higher travel demands were likely caused by Christmas visits to or by nearby residents. The Metropolitan Museum of Art and American Museum of Natural History have the negative Event Index. They were closed on Christmas day and thus less travel demands were observed.

*5.2. Spatiotemporal Events in 2012*

For event detection for 2012, we find the optimal number of sensors was eight, when the validation dataset has the smallest *RMSE* (Figure 6b). Figure 11 and Table 2 show the locations of these eight sensors for 2012. Similar to 2009, Penn Station was the most important sensor location, meaning it indicates the most correlations among taxi travel demands. Similar to 2009, LGA taxi pick up area was also identified as a key location to indicate NYC taxi travel demands. In addition, the JFK airport terminal 4 departure area was found to be a key location. Other than these two airports, all other key locations are all in Manhattan.

**Table 2.** Locations of the 8 sensors for 2012.

| Sensor Number | Location |
| --- | --- |
| 1 | Penn Station |
| 2 | Park Ave and E 53rd St. |
| 3 | 9th Ave and W 16th St—Google Building |
| 4 | JFK Terminal 4 departure |
| 5 | Broadway and W 66th |
| 6 | 11th Ave and W 39th St—Lincoln Tunnel Manhattan end |
| 7 | LGA taxi pick-up area |
| 8 | 2nd Ave and E 52nd St. |

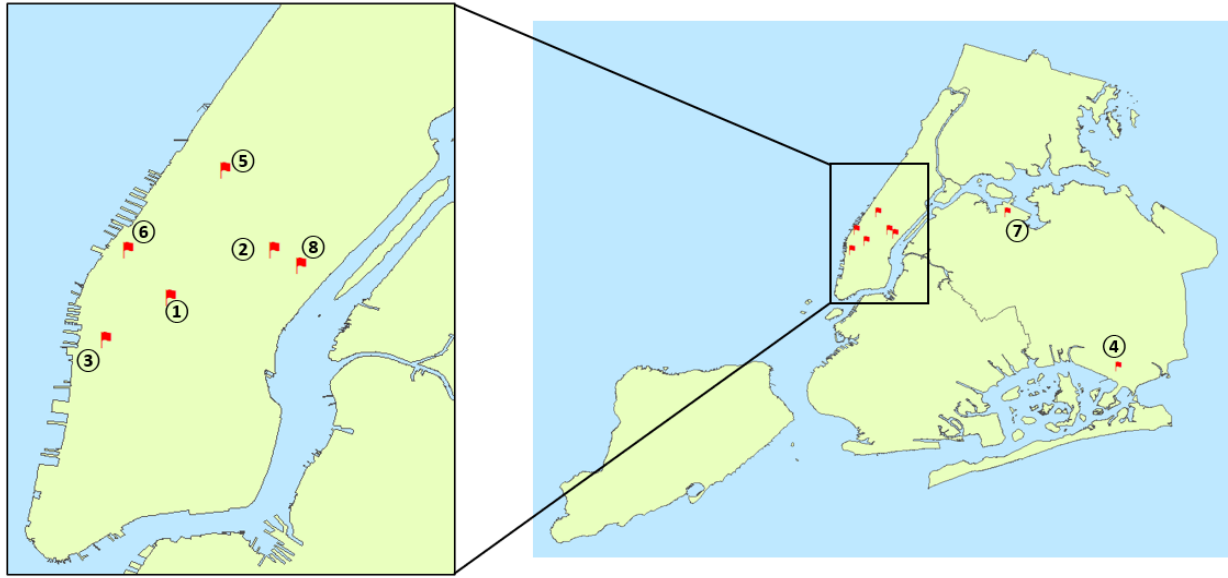

**Figure 11.** Locations of 8 sensors in NYC for 2012 event detection.

Figure 12 shows the everyday *RMSE* over 2012. The day with the largest *RMSE* is Thanksgiving Day (23rd November). The second largest *RMSE* happened on the day when Hurricane Sandy hit NYC (29 October). The third largest *RMSE* appeared on the weekend for St. Patrick's Day (18 March). The day with fourth largest *RMSE* is New Year's Eve (31 December).

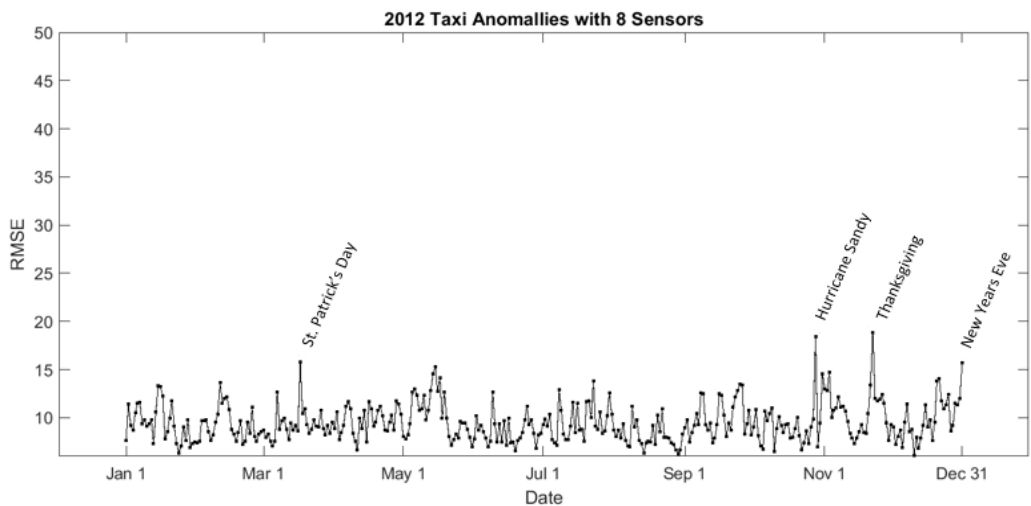

**Figure 12.** Daily *RMSE* showing the difference between the observed travel demand and the simulated uneventful scenario for 2012.

Thanksgiving Day in 2012 also experienced a high *RMSE*, which resulted from several locations with a significantly higher travel demand than estimated. Figure 13a shows the Event Index distribution for Thanksgiving Day. Locations with more travel demand than estimated were gathered in Midtown Manhattan, around Fifth Avenue and Times Square. Most of Downtown Manhattan appeared in blue, meaning a lesser travel demand was observed than estimated. Figure 13b shows the Event Index distribution for the day when Hurricane Sandy hit NYC (29 October). Most of Midtown Manhattan exhibited less travel demands. Midtown is where most companies were located. Many companies were closed and allowed work from home on that day. This means that travel demand for Midtown was less than estimated. The Upper

East Side, which is mostly a residential area, had a higher travel demand than estimated. Higher travel demand was also seen near the NYU Langone Health Hospital and other residential areas. As many subways were closed due to potential flooding, more people chose taxi as the substitute travel mode.

Figure 13c shows the Event Index distribution for St. Patrick's Day. Fifth Avenue is in blue and Madison Avenue, and the next avenue parallel to the Fifth Avenue is in red. This happened because Fifth Avenue was closed for the parade and thus taxi pick-up or drop-off activities happened around the parade route. A higher travel demand was also observed near Times Square and the surrounding commercial areas. Figure 13d shows the Event Index distribution for New Year's Eve. Areas near Times Square exhibited lower than expected travel demand, likely as the area was closed to traffic during the celebration events. Areas surrounding the Times Square had a higher than estimated travel demand, as people who traveled to/from Times Square had to start or end their trips there. In addition, more travel demand was found near the Downtown Financial District and Battery Park, where the ferry to the Statue of Liberty departs.

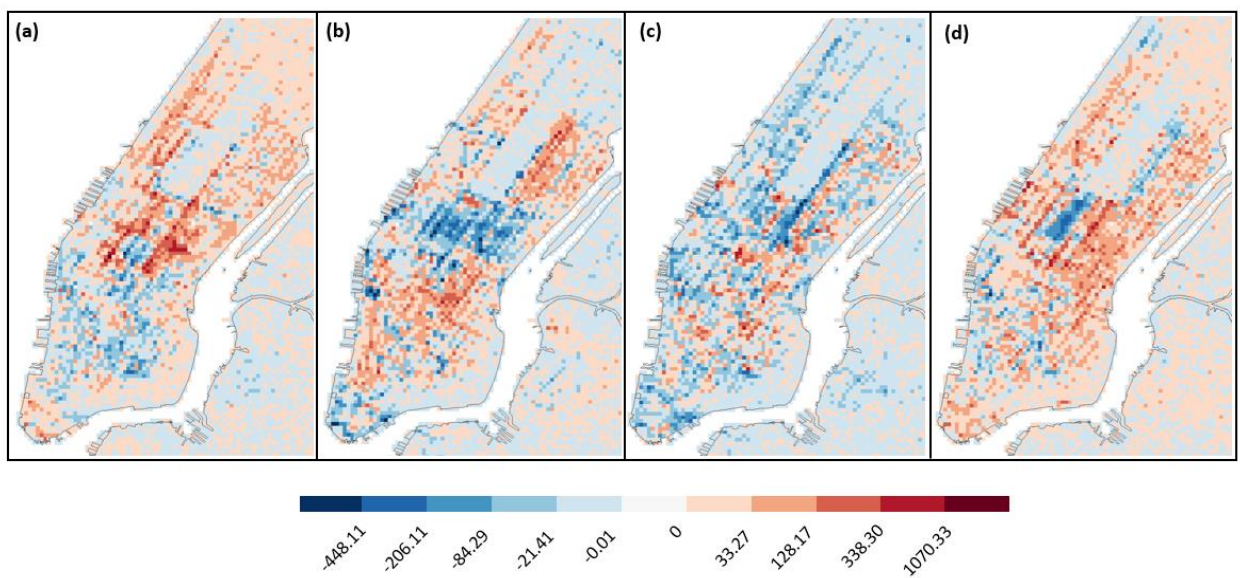

**Figure 13.** Event Index of Manhattan area in 2012 for (**a**) 23 November (Thanksgiving Day), (**b**) 29 October (Hurricane Sandy), (**c**) 18 March (St. Patrick's Day), and (**d**) 31 December (New Year's Eve).

## 6. Conclusions

This paper presents a new method based on the Discrete Empirical Interpolation Method (DEIM) for event detection using point-based human mobility data. This method can first identify the dominant locations in the study area and then simulate an uneventful scenario based only on the limited observation data from the previously identified key locations. Spatiotemporal events are detected by comparing the discrepancy between the observed actual data and the simulated uneventful scenario. Since this method is based on an unsupervised method, it does not require any prior knowledge of the study area or the time window of interest. In addition, this method requires a lower data preparation. Because the simulation process is based on each discrete temporal unit, it does not require chronologically ordered data, which can reduce pre-processing time. The significance of this work extends beyond its technical contributions and can benefit multiple areas, such as urban planning, transportation development, and emergency management. By identifying important locations, it provides the possible solutions for resource allocation and situation awareness. In addition, the real-time application of our method can be used as a powerful tool by the relevant authorities. Understanding the dynamics of events in urban areas is essential for transportation infrastructure optimization, resource allocation, and emergency preparedness.

The proposed novel event detection method demonstrates the feasibility of using the optimal sensor-based simulation method for spatiotemporal event detection. In this paper, locations of sensors are determined by principal components. Since the simulation process is based on observation data at those sensors' locations, events taking place at the sensors' locations cannot be detected. Future research can explore different methods to determine the sensors' locations, such as machine learning, artificial intelligence (AI), and other optimization algorithms. Exploring these alternative methods could provide valuable insights into optimizing sensor deployment for event detection in urban systems. In addition, future research can further utilize this method with other data types, such as social media data or cell phone data. In this work, spatial units and temporal units are arbitrarily determined. In this paper, we used daily travel records in New York City as an illustration, but this method is not limited to this temporal or spatial scale. The temporal resolution is based on the processed dataset for model training and validation. In other words, if the input dataset is processed for hourly travel records, this method can detect anomalies or temporal variability at the hourly scale. Future research can further explore different spatial and temporal unit combinations for practical applications.

**Author Contributions:** Conceptualization, Yuqin Jiang and Andrey A. Popov; methodology, Yuqin Jiang and Andrey A. Popov; analysis and validation, Yuqin Jiang, Andrey A. Popov, Zhenlong Li, and Michael E. Hodgson; writing—original draft preparation, Yuqin Jiang; writing—review and editing, Andrey A. Popov, Zhenlong Li, Michael E. Hodgson, and Binghu Huang; funding acquisition and supervision: Zhenlong Li. All authors have read and agreed to the published version of the manuscript.

**Funding:** The research was in part supported by NSF (2028791) and the University of South Carolina ASPIRE program (135400-20-54176). The funders had no role in study design, data collection and analysis, decision to publish, or preparation of the manuscript.

**Data Availability Statement:** The New York City taxi data used in this study are publicly available at https://www1.nyc.gov/site/tlc/about/tlc-trip-record-data.page (accessed on 20 April 2024).

**Conflicts of Interest:** The authors declare no conflicts of interest.

**References**

1. Peng, C.; Jin, X.; Wong, K.-C.; Shi, M.; Liò, P. Collective Human Mobility Pattern from Taxi Trips in Urban Area. *PLoS ONE* **2012**, *7*, e34487.
2. Yuan, J.; Zheng, Y.; Xie, X. Discovering Regions of Different Functions in a City Using Human Mobility and POIs. In Proceedings of the 18th ACM SIGKDD International Conference on Knowledge Discovery and Data Mining, Beijing China, 12–16 August 2012; pp. 186–194.
3. Barbosa, H.; Barthelemy, M.; Ghoshal, G.; James, C.R.; Lenormand, M.; Louail, T.; Menezes, R.; Ramasco, J.J.; Simini, F.; Tomasini, M. Human mobility: Models and applications. *Phys. Rep.* **2018**, *734*, 1–74. https://doi.org/10.1016/j.physrep.2018.01.001.
4. Isaacman, S.; Becker, R.; Cáceres, R.; Martonosi, M.; Rowland, J.; Varshavsky, A.; Willinger, W. Human mobility modeling at metropolitan scales. In *MobiSys'12: The 10th International Conference on Mobile Systems, Applications, and Services, Low Wood Bay, Lake District, UK, June 25–29, 2012*; ACM: New York, NY, USA, 2012; pp. 239–252.
5. Chen, C.; Ma, J.; Susilo, Y.; Liu, Y.; Wang, M. The promises of big data and small data for travel behavior (aka human mobility) analysis. *Transp. Res. Part C Emerg. Technol.* **2016**, *68*, 285–299. https://doi.org/10.1016/j.trc.2016.04.005.
6. Huang, Z.; Ling, X.; Wang, P.; Zhang, F.; Mao, Y.; Lin, T.; Wang, F.-Y. Modeling real-time human mobility based on mobile phone and transportation data fusion. *Transp. Res. Part C Emerg. Technol.* **2018**, *96*, 251–269. https://doi.org/10.1016/j.trc.2018.09.016.
7. Jiang, Y.; Li, Z.; Cutter, S.L. Social Network, Activity Space, Sentiment, and Evacuation: What Can Social Media Tell Us? *Ann. Am. Assoc. Geogr.* **2019**, *109*, 1795–1810.
8. Martín, Y.; Li, Z.; Cutter, S.L. Leveraging Twitter to gauge evacuation compliance: Spatiotemporal analysis of Hurricane Matthew. *PLoS ONE* **2017**, *12*, e0181701. https://doi.org/10.1371/journal.pone.0181701.
9. Huang, X.; Lu, J.; Gao, S.; Wang, S.; Liu, Z.; Wei, H. Staying at Home Is a Privilege: Evidence from Fine-Grained Mobile Phone Location Data in the United States during the COVID-19 Pandemic. *Ann. Assoc. Am. Geogr.* **2021**, *112*, 286–305. https://doi.org/10.1080/24694452.2021.1904819.
10. Lu, X.; Bengtsson, L.; Holme, P. Predictability of population displacement after the 2010 Haiti earthquake. *Proc. Natl. Acad. Sci. USA* **2012**, *109*, 11576–11581. https://doi.org/10.1073/pnas.1203882109.
11. Song, C.; Qu, Z.; Blumm, N.; Barabási, A.-L. Limits of Predictability in Human Mobility. *Science* **2010**, *327*, 1018–1021. https://doi.org/10.1126/science.1177170.
12. Huang, X.; Li, Z.; Jiang, Y.; Li, X.; Porter, D. Twitter reveals human mobility dynamics during the COVID-19 pandemic. *PLoS ONE* **2020**, *15*, e0241957. https://doi.org/10.1371/journal.pone.0241957.
13. Jiang, Y.; Huang, X.; Li, Z. Spatiotemporal Patterns of Human Mobility and Its Association with Land Use Types during COVID-19 in New York City. *ISPRS Int. J. Geo-Inf.* **2021**, *10*, 344. https://doi.org/10.3390/ijgi10050344.
14. Jiang, Y.; Guo, D.; Li, Z.; Hodgson, M.E. A novel big data approach to measure and visualize urban accessibility. *Comput. Urban Sci.* **2021**, *1*, 10. https://doi.org/10.1007/s43762-021-00010-1.
15. Zhu, X.; Guo, D. Urban event detection with big data of taxi OD trips: A time series decomposition approach. *Trans. GIS* **2017**, *21*, 560–574. https://doi.org/10.1111/tgis.12288.

16. Giannotti, F.; Nanni, M.; Pedreschi, D.; Pinelli, F.; Renso, C.; Rinzivillo, S.; Trasarti, R. Unveiling the complexity of human mobility by querying and mining massive trajectory data. *VLDB J.* **2011**, *20*, 695–719. https://doi.org/10.1007/s00778-011-0244-8.
17. Pan, B.; Zheng, Y.; Wilkie, D.; Shahabi, C. Crowd Sensing of Traffic Anomalies Based on Human Mobility and Social Media. In Proceedings of the 21st ACM SIGSPATIAL International Conference on Advances in Geographic Information Systems, Orlando, FL, USA, 5–8 November 2013; pp. 344–353.
18. Khezerlou, A.V.; Zhou, X.; Li, L.; Shafiq, Z.; Liu, A.X.; Zhang, F. A Traffic Flow Approach to Early Detection of Gathering Events: Comprehensive Results. *ACM Trans. Intell. Syst. Technol. (TIST)* **2017**, *8*, 1–24.
19. Piciarelli, C.; Micheloni, C.; Foresti, G.L. Trajectory-Based Anomalous Event Detection. *IEEE Trans. Circuits Syst. Video Technol.* **2008**, *18*, 1544–1554.
20. Wu, H.; Sun, W.; Zheng, B. A Fast Trajectory Outlier Detection Approach via Driving Behavior Modeling. In Proceedings of the 2017 ACM on Conference on Information and Knowledge Management, Singapore, 6–10 November 2017; pp. 837–846.
21. Li, X.; Li, Z.; Han, J.; Lee, J.-G. Temporal Outlier Detection in Vehicle Traffic Data. In Proceedings of 2009 IEEE 25th International Conference on Data Engineering, Shanghai, China, 29 March–2 April 2009; pp. 1319–1322.
22. Zheng, Y.; Zhang, H.; Yu, Y. Detecting Collective Anomalies from Multiple Spatio-Temporal Datasets across Different Domains. In Proceedings of the 23rd SIGSPATIAL International Conference on Advances in Geographic Information Systems, Bellevue, WA, USA, 3–6 November 2015; pp. 1–10.
23. Dhiman, A.; Toshniwal, D. An Approximate Model for Event Detection from Twitter Data. *IEEE Access* **2020**, *8*, 122168–122184. https://doi.org/10.1109/access.2020.3007004.
24. Weng, J.; Lee, B.-S. Event Detection in Twitter. In Proceedings of the International AAAI Conference on Web and Social Media, Barcelona, Spain, 17–21 July 2011; Volume 5.
25. Zhou, X.; Chen, L. Event detection over twitter social media streams. *VLDB J.* **2014**, *23*, 381–400. https://doi.org/10.1007/s00778-013-0320-3.
26. Dobra, A.; Williams, N.E.; Eagle, N. Spatiotemporal Detection of Unusual Human Population Behavior Using Mobile Phone Data. *PLoS ONE* **2015**, *10*, e0120449. https://doi.org/10.1371/journal.pone.0120449.
27. Traag, V.A.; Browet, A.; Calabrese, F.; Morlot, F. Social Event Detection in Massive Mobile Phone Data Using Probabilistic Location Inference. In Proceedings of the 2011 IEEE Third International Conference on Privacy, Security, Risk and Trust and 2011 IEEE Third International Conference on Social Computing, Boston, MA, USA, 9–11 October 2011; pp. 625–628.
28. Adam, A.; Rivlin, E.; Shimshoni, I.; Reinitz, D. Robust Real-Time Unusual Event Detection using Multiple Fixed-Location Monitors. *IEEE Trans. Pattern Anal. Mach. Intell.* **2008**, *30*, 555–560. https://doi.org/10.1109/tpami.2007.70825.
29. Lu, C.; Shi, J.; Jia, J. Abnormal Event Detection at 150 Fps in Matlab. In Proceedings of the IEEE International Conference on Computer Vision, Sydney, Australia, 1–8 December 2013; pp. 2720–2727.
30. Oh, S.; Hoogs, A.; Perera, A.; Cuntoor, N.; Chen, C.-C.; Lee, J.T.; Mukherjee, S.; Aggarwal, J.K.; Lee, H.; Davis, L.; et al. A large-scale benchmark dataset for event recognition in surveillance video. In Proceedings of the 2011 IEEE Conference on Computer Vision and Pattern Recognition (CVPR), Washington, DC, USA, 20–25 June; pp. 3153–3160.
31. Wan, S.; Xu, X.; Wang, T.; Gu, Z. An Intelligent Video Analysis Method for Abnormal Event Detection in Intelligent Transportation Systems. *IEEE Trans. Intell. Transp. Syst.* **2020**, *22*, 4487–4495. https://doi.org/10.1109/tits.2020.3017505.
32. Ştefănescu, R.; Sandu, A.; Navon, I. POD/DEIM reduced-order strategies for efficient four dimensional variational data assimilation. *J. Comput. Phys.* **2015**, *295*, 569–595. https://doi.org/10.1016/j.jcp.2015.04.030.
33. Brunton, S.L.; Kutz, J.N. *Data-Driven Science and Engineering: Machine Learning, Dynamical Systems, and Control*; Cambridge University Press: Cambridge, UK, 2019; ISBN 1-108-38658-X.
34. Upadhyaya, B.R.; Li, F. Optimal sensor placement strategy for anomaly detection and isolation. In Proceedings of the 2011 Future of Instrumentation International Workshop (FIIW), Oak Ridge, TN, USA, 7–8 November 2011; pp. 95–98.
35. Liu, S.; Auckenthaler, P. Optimal sensor placement for event detection and source identification in water distribution networks. *J. Water Supply Res. Technol.* **2014**, *63*, 51–57. https://doi.org/10.2166/aqua.2013.106.
36. Jayaraman, B.; Al Mamun, S.M.A.; Lu, C. Interplay of Sensor Quantity, Placement and System Dimension in POD-Based Sparse Reconstruction of Fluid Flows. *Fluids* **2019**, *4*, 109. https://doi.org/10.3390/fluids4020109.
37. Abdelhaq, H.; Sengstock, C.; Gertz, M. Eventweet: Online Localized Event Detection from Twitter. *Proc. VLDB Endow.* **2013**, *6*, 1326–1329.
38. Costa, D.G.; Duran-Faundez, C.; Andrade, D.C.; Rocha-Junior, J.B.; Peixoto, J.P.J. TwitterSensing: An Event-Based Approach for Wireless Sensor Networks Optimization Exploiting Social Media in Smart City Applications. *Sensors* **2018**, *18*, 1080. https://doi.org/10.3390/s18041080.
39. Hu, J.; Wang, Y.; Li, P. Online city-scale hyper-local event detection via analysis of social media and human mobility. In Proceedings of the 2017 IEEE International Conference on Big Data (Big Data), Boston, MA, USA, 11–14 December 2017; pp. 626–635.
40. Zhang, W.; Qi, G.; Pan, G.; Lu, H.; Li, S.; Wu, Z. City-Scale Social Event Detection and Evaluation with Taxi Traces. *ACM Trans. Intell. Syst. Technol. (TIST)* **2015**, *6*, 1–20.

41. Zhang, S.; Tang, J.; Wang, H.; Wang, Y. Enhancing Traffic Incident Detection by Using Spatial Point Pattern Analysis on Social Media. *Transp. Res. Rec. J. Transp. Res. Board* **2015**, *2528*, 69–77. https://doi.org/10.3141/2528-08.
42. Jiang, W.; Wang, Y.; Tsou, M.-H.; Fu, X. Using Social Media to Detect Outdoor Air Pollution and Monitor Air Quality Index (AQI): A Geo-Targeted Spatiotemporal Analysis Framework with Sina Weibo (Chinese Twitter). *PLoS ONE* **2015**, *10*, e0141185. https://doi.org/10.1371/journal.pone.0141185.
43. Resch, B.; Usländer, F.; Havas, C. Combining machine-learning topic models and spatiotemporal analysis of social media data for disaster footprint and damage assessment. *Cartogr. Geogr. Inf. Sci.* **2018**, *45*, 362–376. https://doi.org/10.1080/15230406.2017.1356242.
44. Zhang, F.; Li, Z.; Li, N.; Fang, D. Assessment of urban human mobility perturbation under extreme weather events: A case study in Nanjing, China. *Sustain. Cities Soc.* **2019**, *50*, 101671. https://doi.org/10.1016/j.scs.2019.101671.
45. Gao, Y.; Wang, S.; Padmanabhan, A.; Yin, J.; Cao, G. Mapping spatiotemporal patterns of events using social media: A case study of influenza trends. *Int. J. Geogr. Inf. Sci.* **2018**, *32*, 425–449. https://doi.org/10.1080/13658816.2017.1406943.
46. Wang, J.; Zhao, L.; Ye, Y.; Zhang, Y. Adverse event detection by integrating twitter data and VAERS. *J. Biomed. Semant.* **2018**, *9*, 19. https://doi.org/10.1186/s13326-018-0184-y.
47. Jiang, Y.; Li, Z.; Cutter, S.L. Social distance integrated gravity model for evacuation destination choice. *Int. J. Digit. Earth* **2021**, *14*, 1004–1018. https://doi.org/10.1080/17538947.2021.1915396.
48. Sakaki, T.; Okazaki, M.; Matsuo, Y. Earthquake Shakes Twitter Users: Real-Time Event Detection by Social Sensors. In Proceedings of the 19th International Conference on World Wide Web 2010, Raleigh, NC, USA, 26–30 April 2010; pp. 851–860.
49. Wang, Y.; Taylor, J.E. Coupling sentiment and human mobility in natural disasters: A Twitter-based study of the 2014 South Napa Earthquake. *Nat. Hazards* **2018**, *92*, 907–925. https://doi.org/10.1007/s11069-018-3231-1.
50. Yu, M.; Huang, Q.; Qin, H.; Scheele, C.; Yang, C. Deep Learning for Real-Time Social Media Text Classification for Situation Awareness–Using Hurricanes Sandy, Harvey, and Irma as Case Studies. *Int. J. Digit. Earth* **2019**, *12*, 1230–1247.
51. Zhang, C.; Zhou, G.; Yuan, Q.; Zhuang, H.; Zheng, Y.; Kaplan, L.; Wang, S.; Han, J. Geoburst: Real-Time Local Event Detection in Geo-Tagged Tweet Streams. In Proceedings of the 39th International ACM SIGIR conference on Research and Development in Information Retrieval, Pisa, Italy, 17–21 July 2016; pp. 513–522.
52. Andrienko, G.; Andrienko, N.; Hurter, C.; Rinzivillo, S.; Wrobel, S. From movement tracks through events to places: Extracting and characterizing significant places from mobility data. In Proceedings of the 2011 IEEE Conference on Visual Analytics Science and Technology (VAST), Providence, RI, USA, 23–28 October 2011; pp. 161–170.
53. Cui, J.; Liu, F.; Janssens, D.; An, S.; Wets, G.; Cools, M. Detecting urban road network accessibility problems using taxi GPS data. *J. Transp. Geogr.* **2016**, *51*, 147–157. https://doi.org/10.1016/j.jtrangeo.2015.12.007.
54. Ying, J.J.-C.; Lee, W.-C.; Tseng, V.S. Mining geographic-temporal-semantic patterns in trajectories for location prediction. *ACM Trans. Intell. Syst. Technol.* **2014**, *5*, 1–33. https://doi.org/10.1145/2542182.2542184.
55. Jahnke, M.; Ding, L.; Karja, K.; Wang, S. Identifying Origin/Destination Hotspots in Floating Car Data for Visual Analysis of Traveling Behavior. In *Progress in Location-Based Services 2016*; Springer: Berlin/Heidelberg, Germany, 2017; pp. 253–269.
56. Kaiser, M.S.; Lwin, K.T.; Mahmud, M.; Hajializadeh, D.; Chaipimonplin, T.; Sarhan, A.; Hossain, M.A. Advances in Crowd Analysis for Urban Applications Through Urban Event Detection. *IEEE Trans. Intell. Transp. Syst.* **2017**, *19*, 3092–3112. https://doi.org/10.1109/tits.2017.2771746.
57. Fekih, M.; Bellemans, T.; Smoreda, Z.; Bonnel, P.; Furno, A.; Galland, S. A data-driven approach for origin–destination matrix construction from cellular network signalling data: A case study of Lyon region (France). *Transportation* **2020**, *48*, 1671–1702. https://doi.org/10.1007/s11116-020-10108-w.
58. Abd-Alrazaq, A.; Alhuwail, D.; Househ, M.; Hamdi, M.; Shah, Z. Top Concerns of Tweeters During the COVID-19 Pandemic: Infoveillance Study. *J. Med. Internet Res.* **2020**, *22*, e19016. https://doi.org/10.2196/19016.
59. Chen, X.; Wang, S.; Tang, Y.; Hao, T. A bibliometric analysis of event detection in social media. *Online Inf. Rev.* **2019**, *43*, 29–52. https://doi.org/10.1108/oir-03-2018-0068.
60. Jelodar, H.; Wang, Y.; Yuan, C.; Feng, X.; Li, Y.; Zhao, L. Latent Dirichlet Allocation (LDA) and Topic modeling: Models, applications, a survey. *Multimed. Tools Appl.* **2019**, *78*, 15169–15211.
61. Alqhtani, S.M.; Luo, S.; Regan, B. Fusing Text and Image for Event Detection in Twitter. *arXiv* **2015**, arXiv:1503.03920.
62. Huang, X.; Li, Z.; Wang, C.; Ning, H. Identifying disaster related social media for rapid response: A visual-textual fused CNN architecture. *Int. J. Digit. Earth* **2019**, *13*, 1017–1039. https://doi.org/10.1080/17538947.2019.1633425.
63. Jiang, Y.; Li, Z.; Ye, X. Understanding demographic and socioeconomic biases of geotagged Twitter users at the county level. *Cartogr. Geogr. Inf. Sci.* **2019**, *46*, 228–242. https://doi.org/10.1080/15230406.2018.1434834.
64. Malik, M.; Lamba, H.; Nakos, C.; Pfeffer, J. Population Bias in Geotagged Tweets. *People* **2015**, *1*, 3–759.
65. Mellon, J.; Prosser, C. Twitter and Facebook are not representative of the general population: Political attitudes and demographics of British social media users. *Res. Politics* **2017**, *4*, 2053168017720008. https://doi.org/10.1177/2053168017720008.

66. Toch, E.; Lerner, B.; Ben-Zion, E.; Ben-Gal, I. Analyzing large-scale human mobility data: A survey of machine learning methods and applications. *Knowl. Inf. Syst.* **2019**, *58*, 501–523. https://doi.org/10.1007/s10115-018-1186-x.
67. Zhou, S.; Shen, W.; Zeng, D.; Zhang, Z. Unusual event detection in crowded scenes by trajectory analysis. In Proceedings of the ICASSP 2015 IEEE International Conference on Acoustics, Speech and Signal Processing (ICASSP), South Brisbane, QLD, Australia, 19–24 April 2015; pp. 1300–1304.
68. Tang, J.; Liu, F.; Wang, Y.; Wang, H. Uncovering urban human mobility from large scale taxi GPS data. *Phys. A Stat. Mech. Appl.* **2015**, *438*, 140–153. https://doi.org/10.1016/j.physa.2015.06.032.
69. Luo, F.; Cao, G.; Mulligan, K.; Li, X. Explore spatiotemporal and demographic characteristics of human mobility via Twitter: A case study of Chicago. *Appl. Geogr.* **2016**, *70*, 11–25. https://doi.org/10.1016/j.apgeog.2016.03.001.
70. Tang, J.; Zhang, S.; Chen, X.; Liu, F.; Zou, Y. Taxi trips distribution modeling based on Entropy-Maximizing theory: A case study in Harbin city—China. *Phys. A Stat. Mech. Appl.* **2018**, *493*, 430–443. https://doi.org/10.1016/j.physa.2017.11.114.
71. Qin, S.-M.; Verkasalo, H.; Mohtaschemi, M.; Hartonen, T.; Alava, M. Patterns, Entropy, and Predictability of Human Mobility and Life. *PLoS ONE* **2012**, *7*, e51353. https://doi.org/10.1371/journal.pone.0051353.
72. Kulldorff, M. A Spatial Scan Statistic. *Commun. Stat. Theory Methods* **1997**, *26*, 1481–1496.
73. Austwick, M.Z.; O'brien, O.; Strano, E.; Viana, M. The Structure of Spatial Networks and Communities in Bicycle Sharing Systems. *PLoS ONE* **2013**, *8*, e74685. https://doi.org/10.1371/journal.pone.0074685.
74. Alfieri, L.; Thielen, J.; Pappenberger, F. Ensemble hydro-meteorological simulation for flash flood early detection in southern Switzerland. *J. Hydrol.* **2012**, *424–425*, 143–153. https://doi.org/10.1016/j.jhydrol.2011.12.038.
75. Younis, J.; Anquetin, S.; Thielen, J. The benefit of high-resolution operational weather forecasts for flash flood warning. *Hydrol. Earth Syst. Sci.* **2008**, *12*, 1039–1051. https://doi.org/10.5194/hess-12-1039-2008.
76. Kalnay, E. *Atmospheric Modeling, Data Assimilation and Predictability*; Cambridge University Press: Cambridge, UK, 2003; ISBN 0-521-79179-0.
77. Evensen, G. *Data Assimilation: The Ensemble Kalman Filter*; Springer Science & Business Media: Berlin, Germany, 2009; ISBN 3-642-03711-9.
78. Shead, T.; Tezaur, I.; Davis IV, W.; Carlson, M.; Dunlavy, D.; Parish, E.; Blonigan, P.; Tencer, J.; Rizzi, F.; Kolla, H. A Novel In Situ Machine Learning Framework for Intelligent Data Capture and Event Detection. In *Machine Learning and Its Application to Reacting Flows: ML and Combustion*; Springer International Publishing: Cham, Switzerland, 2023; pp. 53–87.
79. Chawla, S.; Zheng, Y.; Hu, J. Inferring the Root Cause in Road Traffic Anomalies. In Proceedings of the 2012 IEEE 12th International Conference on Data Mining, Brussels, Belgium, 10–13 December 2012; pp. 141–150.
80. Yang, S.; Kalpakis, K.; Biem, A. Detecting Road Traffic Events by Coupling Multiple Timeseries with a Nonparametric Bayesian Method. *IEEE Trans. Intell. Transp. Syst.* **2014**, *15*, 1936–1946. https://doi.org/10.1109/tits.2014.2305334.
81. Yang, S.; Zhou, W. Anomaly Detection on Collective Moving Patterns: Manifold Learning Based Analysis of Traffic Streams. In Proceedings of the 2011 IEEE Third International Conference on Privacy, Security, Risk and Trust (PASSAT)/2011 IEEE Third International Conference on Social Computing (SocialCom), Boston, MA, USA, 9–11 October; pp. 704–707.
82. Abdi, H.; Williams, L.J. Principal Component Analysis. *Wiley Interdiscip. Rev. Comput. Stat.* **2010**, *2*, 433–459.
83. Ringnér, M. What Is Principal Component Analysis? *Nat. Biotechnol.* **2008**, *26*, 303–304.
84. Wold, S.; Esbensen, K.; Geladi, P. Principal Component Analysis. *Chemom. Intell. Lab. Syst.* **1987**, *2*, 37–52.
85. Good, P.I.; Hardin, J.W. *Common Errors in Statistics (and How to Avoid Them)*; John Wiley & Sons: Hoboken, NJ, USA, 2012; ISBN 1-118-36011-7.
86. Jung, S.; Marron, J.S. PCA consistency in high dimension, low sample size context. *Ann. Stat.* **2009**, *37*, 4104–4130. https://doi.org/10.1214/09-aos709.
87. R: Jenks Natural Breaks Classification. Available online: https://search.r-project.org/CRAN/refmans/BAMMtools/html/getJenksBreaks.html (accessed on 28 January 2022).
88. Jenks, G.F. The Data Model Concept in Statistical Mapping. *Int. Yearb. Cartogr.* **1967**, *7*, 186–190.
89. Chen, J.; Yang, S.T.; Li, H.W.; Zhang, B.; Lv, J.R. Research on geographical environment unit division based on the method of natural breaks (Jenks). *Int. Arch. Photogramm. Remote Sens. Spat. Inf. Sci.* **2013**, *40*, 47–50. https://doi.org/10.5194/isprsarchives-xl-4-w3-47-2013.
90. McMaster, R. In Memoriam: George f. Jenks (1916–1996). *Cartogr. Geogr. Inf. Syst.* **1997**, *24*, 56–59.